\documentclass[aps, pra, 10pt, twocolumn, superscriptaddress, floatfix,longbibliography,nofootinbib]{revtex4-2}
\usepackage[T1]{fontenc}
\usepackage[utf8]{inputenc}
\usepackage[english]{babel}
\usepackage{braket}
\usepackage[plainpages=false,colorlinks=true,linkcolor=blue, citecolor=blue, urlcolor=blue]{hyperref}
\usepackage[charter]{mathdesign}
\usepackage[caption=false]{subfig}
\usepackage{amsmath,graphicx,bm}
\usepackage{xcolor}
\usepackage{cleveref}

\newcommand\B{\mathbf{B}}
\newcommand\C{\mathbf{C}}

\begin{document}
\title{Liouvillian recursion method for the electronic Green function} 

\author{A. Foley}
\email[Corresponding author: ]{Alexandre.Foley@usherbrooke.ca}
\affiliation{AlgoLab Quantique, Institut Quantique, Universit\'e de Sherbrooke, Sherbrooke, QC, Canada}
\date{\today}

\begin{abstract}
In this paper, we present a framework for the recursion method applied within the Liouvillian formalism, enabling the computation of response functions for a wide range of quantum operators.
Within the presented framework, one can readily compute the electronic Green function, or the equilibrium response function of any operators, hermitian or not.
This is in contrast with most of the previous literature on the Liouvillian recursion method, in which only hermitian operators can be treated.
Additionally, within the presented framework, one can easily extend the recursive method to perform orthonormal polynomial decomposition of correlation functions.
We demonstrate the method by showcasing its application to a temperature dependence study of a $2\times2$ Hubbard model. 
Our results show that the recursion method accurately computes response functions and provides valuable insights into quantum many-body systems. 
While the method has already been successfully applied to study of condensed matter, further efforts are required to develop efficient algorithms for larger and more realistic systems. 
Tensor network methods and other variational techniques hold promise for tackling these challenges. 
Overall, the recursion method offers a flexible approach for investigating strongly correlated system.
\end{abstract}

\maketitle

\section{Introduction}
The recursion method in the Liouvillian formalism has been extensively used for the computation of response functions of Hermitian operators~\cite{viswanath_recursion_1994,mori_continued-fraction_1965,lee_orthogonalization_1982,lee_solutions_1982,lee_derivation_1983,balucani_dynamical_2003}. 
However, it is not the case for response function of non-Hermitian operators, such as the electronic Green function.
Indeed, the typical presentation~\cite{viswanath_recursion_1994,lee_orthogonalization_1982,balucani_dynamical_2003} doesn't allow for such a response function.

While Mori's original derivation~\cite{mori_continued-fraction_1965} of the approach is applicable to the general case, it is rather complicated~\cite{balucani_dynamical_2003,lee_derivation_1983}.
A goal of this work is to derive Mori's equations with modern and transparent mathematical techniques, making the breadth of its applicability clearer.
The computation of electronic Green function is routinely done using the recursion method in the Hamiltonian formalism~\cite{dionne_pyqcm_2023,dagotto_correlated_1994,georges_dynamical_1996,viswanath_recursion_1994}. 
This has some advantages and some drawbacks: the numerical implementation is simple and efficient, but the  approach is only readily applicable in the zero-temperature limit.
In contrast, with the Liouvillian formalism the approach can be carried out at arbitrary temperature, but the superoperator algebra is considerably harder to implement efficiently.
We will show that the recursion method within the Hamiltonian formalism is a special case of the Liouvillian formalism.

In its canonical use~\cite{viswanath_recursion_1994,mori_continued-fraction_1965}, the recursion method computes a response function of a finite system in the form of a continued fraction and, by analysing the structure of this continued fraction, information about the infinite-size limit can be inferred.
This inferred data is often interesting in its own right and from it an approximation of the response function of the infinite system can be reconstructed.
The reconstruction process takes the form of an extrapolation of the continued fraction explicitly computed.
It can also be used in conjunction with quantum cluster methods~\cite{maier_quantum_2005}, sometimes with the structural analysis of the continued fraction~\cite{barnabe-theriault_analyse_2003, premont-foley_reseaux_2020}, but more often without~\cite{dionne_pyqcm_2023,dagotto_correlated_1994,georges_dynamical_1996}.
This paper focuses on the explicit computation of continued fractions, we refer the reader to the monograph by Viswanath and Muller~\cite{viswanath_recursion_1994} for further reading on continued fraction analysis and extrapolation.

This paper is organized as follows: we first explain and derive the method;we then give a few examples of functions that can be computed within this formalism and demonstate its usage with a study in temperature of a $2\times2$ Hubbard model.
Finally, we present our conclusions.

\section{Method}\label{sec:method}
In this section, we first present the notation and basic definitions and then use those to obtain the equations of the recursion method.
We then derive a decomposition of the spectral function and Green function in terms of orthonormal polynomials.
Finally, we explore the relation of that decomposition with the recursion method.

\subsection{Definitions and notation}\label{sec:notation}
We work with natural units $k_B = \hbar = 1$, operators that are not explicitly function of time are understood to be evaluated at time $t=0$ and the Fourier transform of a function $f(t)$ is $f(\omega) = \int_{-\infty}^\infty f(t)e^{i\omega t}dt$.

The Liouvillian recursion method can be understood quite simply in terms of Heisenberg's equation in a superoperator formalism: the Liouvillian superoperator $\mathcal{L}$ is the commutator of some operator with the Hamiltonian.
Heisenberg's equation, in the Liouvillian notation reads as follows:
\begin{equation}
    -i \frac{\partial}{\partial t} \mathbf{O}(t) = \mathcal{L}\mathbf{O}(t) \equiv [\mathbf{H},\mathbf{O}(t)], 
    \label{eq:heis}
\end{equation}
where $\mathbf{O}$ is some operator, $\mathbf{H}$ is the Hamiltonian and $\mathcal{L}\equiv [\mathbf{H},\cdot]$ is the Liouvillian.
We quickly observe that in terms of this superoperator, Heisenberg's equation has the same structure as Schrödinger's equation.
It consequently admits the same formal solution for time evolution~\cite{viswanath_recursion_1994,hatano_finding_2005}:
\begin{equation}
    \mathbf{O}(t) = e^{i \mathcal{L} t} \mathbf{O} \equiv  e^{i \mathbf{H} t} \mathbf{O} e^{-i \mathbf{H} t}\label{eq:timevol}
\end{equation}

With this last observation, there's only one missing ingredient to perform the computation of response functions or their associated Green functions: a scalar product for operators.
While its precise definition is of little importance for the relation we will demonstrate next, the precise choice of scalar product is physically motivated: it determines what sort of response functions we will compute.
Concrete exemples of such scalar products are given in subsection~\ref{subs:scalarprod}.
We denote the scalar product of two operators $\mathbf{B}$ and $\mathbf{C}$ as $(\B|\C)$, then the response function and retarded Green function available to the recursion method are 
\begin{align}
    A_{\B\C}(t) &= (\mathbf{B}|\mathbf{C}(t)) \label[type]{eq:sp_SW}
    \\
    G_{\B\C}(t) &= -i (\mathbf{B}|\mathbf{C}(t))\theta(t) \label{eq:sp_green}
\end{align}
respectively, for any pair of operators.
In Eq.~\eqref{eq:sp_green}, $\theta(t)$ is the Heaviside step function.
We use parentheses to denote the scalar product instead of the more conventional angled brackets to distinguish the operation clearly from the scalar product between states, because practical definition of the operator scalar product can involve that of states.

The recursion method requires that the Liouvillain be hermitian with respect to the scalar product:
\begin{align}
    (\mathbf{C}|\mathcal{L}\mathbf{B}) = (\mathcal{L}\mathbf{C}|\mathbf{B}) \equiv (\mathbf{C}|\mathcal{L}|\mathbf{B}).
    \label{eq:equilibrium}
\end{align}
With that additional property, the scalar product reproduces the usual algebra of Dirac's notation with the Hamiltonian in every way. 
Note that this extra property~\eqref{eq:equilibrium} is readily obtained when the scalar product is the average at thermodynamic equilibrium of some operator.
\subsection{Recursion method for continued fractions}
Using the time evolution identity~\eqref{eq:timevol},we can rewrite Eq.~\eqref{eq:sp_green} as follows:
    \begin{equation}
    \begin{split}
        G_{\mathbf{BC}}(t) &= -i(\mathbf{B}|e^{ i \mathcal{L} t}\mathbf{C})\theta(t) \\
        G_{\mathbf{BC}}(t) &= -i(\mathbf{B}|e^{ i \mathcal{L} t} | \mathbf{C}) \theta(t).
    \end{split}
    \end{equation}
The second line is a consequence of Eq.~\eqref{eq:equilibrium}.
Since it is trivially applicable to any power of the Liouvillian, it also works with any Taylor series and consequently any function of the Liouvillian.
Next, we compute the Fourier transform with respect to time and obtain
\begin{equation}
    G_{\mathbf{B}\mathbf{C}}(\omega) = (\mathbf{B}|\left[{\omega+\mathcal{L}}\right]^{-1}|\mathbf{C}).
\end{equation}
In the particular case $\mathbf{B} = \mathbf{C}$, the problem amounts to computing a single diagonal element of $[\omega+\mathcal{L}]^{-1}$.
This computation is readily accomplished using Lanczos' algorithm~\cite{dionne_pyqcm_2023,dagotto_correlated_1994} on the Liouvillian using $\mathbf{B}$ as the initial operator.
Lanczos' algorithm is an iterative procedure that constructs an orthogonal basis in which the target linear operator is a tridiagonal matrix.
When applied to minus one times the Liouvillian, the algorithm can be expressed by the following recursive relation:
\begin{align}
    \beta_{k+1}\mathbf{f}_{k+1} &= \mathbf{\tilde{f}}_{k+1} = -\mathcal{L}\mathbf{f}_k - \alpha_k
     \mathbf{f}_k - \beta_k \mathbf{f}_{k-1} 
     \label[type]{eq:LLanczos_recursion}
     \\  
     (\mathbf{f}_k|\mathbf{f}_l) &= \delta_{ij}\\
    \alpha_k &= -(\mathbf{f}_k|\mathcal{L}|\mathbf{f}_k) \label[type]{eq:Lanczos_alpha} \\
    \beta_k &= -(\mathbf{f}_k|\mathcal{L}|\mathbf{f}_{k-1}) \label[type]{eq:Lanczos_beta}\\
    \beta_k^2 &= (\tilde{\mathbf{f}}_k|\tilde{\mathbf{f}}_k) \label{eq:beta_practical}\\
    \mathbf{f}_{-1} &= 0 \\
    \mathbf{f}_0 &= \frac{\mathbf{B}}{\sqrt{(\mathbf{B}|\mathbf{B})}}\label{eq:init_f0}
\end{align}
In the basis of $\mathbf{f}_{k}$ operators and after $n$ iterations, the Liouvillian is a tridiagonal matrix:
\begin{equation}
    T = \begin{bmatrix}
        \alpha_0 & \beta_1  & 0       & \dots     &0          \\
        \beta_1  & \alpha_1 & \beta_2 & \ddots    & \vdots    \\
        0        & \beta_2  &\alpha_2 & \ddots    & 0         \\
        \vdots   & \ddots   & \ddots  & \ddots    & \beta_{n} \\
        0        & \dots    &0        & \beta_{n} & \alpha_n  
    \end{bmatrix}.
\end{equation}
Because we used Eqs.~\eqref{eq:init_f0} to initialize the recursion, we obtain the following expression for the Green function:
\begin{equation}
    G_{\mathbf{BB}}(\omega) = (\mathbf{B}|\mathbf{B})\left[\frac{1}{\omega-T}\right]_{(0,0)}
\end{equation}
it can be shown that the $(0,0)$ element of the inverse matrix is a Jacobi continued fraction~\cite{dagotto_correlated_1994,sup_mat} :
\begin{equation}
    G_{\mathbf{BB}}(\omega) = \cfrac{(\mathbf{B}|\mathbf{B})}{\omega- \alpha_0 - \cfrac{\beta_1^2}{\omega - \alpha_1 - \cfrac{\beta_2^2}{\ddots}}}
\end{equation}
Continued fractions are characterized by a number of floors; the number of divisions present in the continued fraction.
With Jacobi continued fractions, the number of floors counts the number of poles the function has on the neighbourhood of the real axis.
A Green’s function with a continuous spectral function necessarily has an infinite number of floors to its continued fraction representation.
Truncating a continued fraction yields rapidly converging approximations for the purpose of computing integrals of the complete continued fraction~\cite{viswanath_recursion_1994,wall_analytic_1948}.
While this method is limited to the direct computation of diagonal elements of the Green function, it doesn't prevent us from computing off-diagonal elements as well, albeit in an indirect manner.
Indeed, by computing the Green function for the operators $\mathbf{B}+\mathbf{C}$ and $\mathbf{B}+i\mathbf{C}$ as well as for $\mathbf{B}$ and $\mathbf{C}$, Lanczos' algorithm yields a continued fraction for
\begin{equation}
\begin{split}
    G^r_{\mathbf{BC}}(\omega) &= ({\B+\C}|\frac{1}{\omega+\mathcal{L}}|{\B+\C}) \\
    &= G_{\mathbf{BB}}(\omega) + G_{\mathbf{BC}}(\omega) + G_{\mathbf{CB}}(\omega) + G_{\mathbf{CC}}(\omega).
    \label[type]{eq:LCCFR}
\end{split}
\end{equation}
and 
\begin{equation}
\begin{split}
    G^i_{\mathbf{BC}}(\omega) &= (\mathbf{B}+i\mathbf{C}|\frac{1}{\omega+\mathcal{L}}|\B+i\C) \\
    &= G_{\mathbf{BB}}(\omega) + iG_{\mathbf{BC}}(\omega) -i G_{\mathbf{CB}}(\omega) + G_{\mathbf{CC}}(\omega).
    \label[type]{eq:LCCFI}
\end{split}
\end{equation}
From the above, expressions for $G_{\mathbf{CB}}$ and $G_{\mathbf{BC}}$ can be extracted.
We refer to this procedure as {\it linear combination of continued fractions} (LCCF).
In the generic case, the computational cost for off-diagonal Green functions is increased fourfold relative to computing a single diagonal element, but it yields four Green functions which are often all of equal interest.

To reach an exact result, the Lanczos' algorithm has to be run until exhaustion of the Krylov subspace generated by the repeated action of the Liouvillian on the initial operator.
Note that subspace exhaustion has happened as soon as $\beta_{k+1}$ in Eqs.~\ref{eq:LLanczos_recursion} is zero to working precision. 
In the case of electrons, this means $\mathcal{O}( 4^n)$ iterations, where $n$ is the number of single-electron orbitals.
This is often neither practical nor necessary.
Indeed, the continued fraction decomposition reveals a lot of information within a few tens of iterations~\cite{viswanath_recursion_1994}, even in the limit of large $n$. 

\subsection{Orthonormal polynomials decomposition}
Within this Liouvillian framework, it is also possible to compute the decomposition of a response function in a basis of orthonormal polynomials.
Orthonormal-polynomial decompositions of response functions rely, as the name suggests, on orthonormal polynomials~\cite{weisse_kernel_2006}. 
The Liouvillian version of the orthonormal polynomial method has one considerable advantage over the equivalent Hamiltonian approach: Because it can compute contributions to the spectral function of both occupied and unoccupied states at once, it doesn't have to deal with the discontinuity of the Fermi-Dirac distribution in metallic systems at zero temperature. 
This means faster convergence of the polynomial coefficients is expected without resorting to subtle tricks~\cite{wolf_chebyshev_2014}.
Note that Hermiticity of the Liouvillian is unnecessary for this method.
There are many families of orthonormal polynomials and they all share a common set of properties. 
Consider the set of orthonormal polynomials $\{T_n(\omega)\} \forall n \in \mathbb N $.
They define a function space with an inner product 
\begin{equation}
\langle f,g \rangle = \int_a^b f(\omega)g(\omega)W(\omega)d\omega
\label{eq:poly_in_prod}
\end{equation}
with respect to a weight function $W(\omega)$ which is real and positive for all $\omega$ in $[a,b]$ and normalized: $\int_a^b W(\omega)d\omega=1$.
The polynomial basis is orthonormal with respect to this inner product:
\begin{equation}
    \langle T_n,T_m \rangle = \delta_{nm}
    \label[type]{eq:poly_ortho}
\end{equation}
The polynomials in the set are related by a three-term recursion relation:
\begin{equation}
    b_{k+1}T_{k+1}(\omega) = (\omega-a_k)T_k(\omega) - b_kT_{k-1}(\omega)
    \label[type]{eq:poly_recur}
\end{equation}
where the coefficients $a_k$ and $b_k$ are determined by the set of orthonormal polynomials, $T_0(\omega)= 1$ and $T_{-1}(\omega)=0$.
The weight function can be obtained from the coefficients of the three-term recursion relation~\cite{wall_analytic_1948,cuyt_handbook_2008}: 
\begin{equation}
     W(\omega) = -\lim_{\epsilon \rightarrow 0^+} \frac{1}{\pi}\mathrm{Im}\left[\vcenter{\hbox{$\displaystyle \cfrac{1}{\omega+ i\epsilon- a_0 - \cfrac{b_1^2}{\omega + i\epsilon - a_1 - \cfrac{b_2^2}{\ddots}}} $ }} \right].\label{eq:weightcf}
\end{equation}
A function $f(\omega)$ defined on the domain $[a,b]$ can be decomposed on the polynomial basis:
\begin{align}
    f(\omega) &= \sum_n \mu_n T_n(\omega) W(\omega) 
    \label[type]{eq:mu_decomposition}
    \\
    &= \sum_n \nu_n T_n(\omega)
    \label[type]{eq:nu_decomposition}
    \\
    \mu_n &= \int_a^b f(\omega)T_n(\omega)d\omega = \langle \tfrac{f}{W},T_n \rangle
    \label[type]{eq:mu_moments}
    \\
    \nu_n &= \int_a^b f(\omega)T_n(\omega)W(\omega)d\omega = \langle f,T_n \rangle
    \label[type]{eq:nu_moments}
\end{align}
Examples of such polynomials are Chebyshev, Legendre, Hermite and Laguerre polynomials.
The first two are defined on a finite domain, while Laguerre's polynomials are on a semi-infinite domain and Hermite's are defined for all real values.
For all four of those polynomial bases, the weight functions $W(\omega)$ are known in a closed form, e.g. Eq.~\eqref{eq:weightcf} is unnecessary.
Note that some guesswork is needed to adapt the polynomial basis to the a priori unknown bandwidth of the Green function as shown in Refs\cite{weisse_kernel_2006,wolf_chebyshev_2014}.

Along with the identities introduced in \ref{sec:notation}, the equations introduced in this section allow us to devise an algorithm to compute the $\mu$-moments, defined in Eq.~\eqref{eq:mu_moments}, of a response function in the frequency domain.
By performing a Fourier transform on Eq.~\eqref{eq:sp_SW} we obtain:
\begin{equation}
    A_{\mathbf{BC}}(\omega) = (\mathbf{B}|\delta(\omega+\mathcal{L})|\mathbf{C})
\end{equation} 
Then substituting $f(\omega)$ by $A_{\mathbf{BC}}(\omega)$ in Eq.~\eqref{eq:mu_moments} yields:
\begin{equation}
    \begin{split}
        \mu_n &= \int_a^b (\mathbf{B}|\delta(\omega+\mathcal{L})|\mathbf{C})T_n(\omega)d\omega\\
        &= (\mathbf{B}|T_n(-\mathcal{L})|\mathbf{C}) \\
        &= (\mathbf{B}|\mathbf{f}_n),
    \end{split}
\end{equation}
where we defined $\mathbf{f}_n = T_n(-\mathcal{L})\mathbf{C}$.
Multiplying the polynomials' three-term recursion~\eqref{eq:poly_recur} by $\mathbf{C}$, we obtain
\begin{equation}
    \begin{split}
         b_{k+1}T_{k+1}(-\mathcal{L})\mathbf{C} &= (-\mathcal{L}-a_k)T_k(-\mathcal{L})\mathbf{C} - b_kT_{k-1}(-\mathcal{L})\mathbf{C} \\
        \implies  b_{k+1} \mathbf{f}_{k+1} &= (-\mathcal{L}-a_k)\mathbf{f}_k - b_k \mathbf{f}_{k-1}
    \end{split}
\end{equation}
The set of operators $\mathbf{f}_k$ can be efficiently computed to finite order without explicitly evaluating the polynomials of the superoperator.

Unlike Lanczos's algorithm, the formal solution produced by this method is typically an infinite series, independently of system size.
This isn't much of a practical issue, because the Lanczos recursion and the orthonormal polynomial methods explore the same Krylov subspace and extract a similar amount of information at a similar pace. 
Indeed, a unique $N$-pole approximation of the Green function could be reconstructed from $2N$ polynomial moments~\cite{wheeler_modified_1984}.
Note that naive algorithms to compute this $N$-pole approximation from polynomial moments are ill-conditioned~\cite{liu_computation_2021}.
In practice, we must truncate the polynomial decomposition to finite order.
Doing so causes Gibbs oscillations to the reconstructed spectral function~\cite{weisse_kernel_2006}.
This phenomenon can be controlled using kernel smoothing methods~\cite{weisse_kernel_2006} or linear prediction~\cite{wolf_chebyshev_2014,ganahl_efficient_2015}. 
Finally, a striking advantage of the polynomial decomposition, relative to Lanczos' recursion, is that multiple response functions can be computed in a single recursive procedure: 
the operator $\mathbf{B}$ does not intervene in the recursive part of the algorithm.
Consequently, the decomposition of any response function that shares the same $\mathbf{C}$ operator can be computed using the same set of $\mathbf{f}_k$ operators.

This recursive procedure produces a decomposition of the spectral function on the chosen basis, but the Green function can also be obtained from the same coefficients.
The Green function and the spectral function in the frequency domain are related by the Kramers-Kronig relations~\cite{dressel_electrodynamics_2002,fetter_quantum_1971}.
A corollary of these relations is the analytic continuation relations:
\begin{align}
    G_{\B\C}(z) &= \lim_{\epsilon \rightarrow 0^+} \frac{1}{2\pi} \int_{-\infty}^\infty \frac{A_{BC}(\omega)}{z-\omega+i\epsilon} d\omega,
    \label[type]{eq:Stieltjes_SW}
    \\
    A_{\B\C}(\omega) &= -\lim_{\epsilon \rightarrow 0^+} \frac{1}{2} \mathrm{Im}\left[ G(\omega+i\epsilon)\right].
    \label[type]{eq:Stieltjes_inversion}
\end{align}
We observe in Eq.~\eqref{eq:Stieltjes_SW} that the Green function is the Stieltjes transform of the spectral function.
The theory relative to the Stieltjes transform~\cite{cuyt_handbook_2008,zayed_handbook_1996,wall_analytic_1948} is deeply intertwined with that of orthonormal polynomials.
Given a $\mu$-decomposition of $A_{BC}(\omega)$, the solution to Eq.~\eqref{eq:Stieltjes_SW} takes the form~\cite{groux_sur_2007,sherman_numerators_1933}:
\begin{equation}
    G_{\B\C}(z) = \sum_n \mu_n \left( U_n(z) + T_n(z)S_W(z) \right)
\end{equation}
where $S_W(z)$ is the Stieltjes transform of the weight function $W(\omega)$ and the $U_n(z)$ are the secondary polynomials, defined in Eqs.~(\ref{eq:secondary_recur}-\ref{eq:secondary_recur_init1}), of the same weight function.
If the function $S_W(z)$ is not known analytically, it can be computed using a continued fraction involving the coefficients of the three-term recursion:
\begin{equation}
    S_W(z) = {\hbox{$\displaystyle  \cfrac{1}{z- a_0 - \cfrac{b_1^2}{z - a_1 - \cfrac{b_2^2}{\ddots}}} $ }}.
\end{equation} 
The secondary polynomials are defined by the following recursion and initial conditions:
\begin{align}
    b_{k+1}U_{k+1}(z) &= (z-a_k)U_k(z) - b_kU_{k-1}(z)
    \label{eq:secondary_recur}\\
    U_0(z) &= 0 \label{eq:secondary_recur_init0}\\
    U_1(z) &= \frac{-T_1(z)}{z-a_0} \label{eq:secondary_recur_init1}
\end{align}
The initial conditions for the $U_i$ polynomials are fixed by the requirement that $ G_{\mathbf{BC}}(z) \simeq {(\mathbf{B}|\mathbf{C})}/{z} $ when $|z| \gg 1$.
In other words, all terms in $z^n$ $(n\geq 0)$ must cancel at high frequency.

\subsection{On the relation between continued fractions and orthonormal polynomials}
One quickly notices that the Lanczos algorithm and the orthonormal polynomial method rely on basically identical equations. 
In fact, Lanczos' procedure amounts to generating a special-purpose polynomial basis such that $\mu_k = (\mathbf{B}|\mathbf{B})\delta_{k,0}$. 
Indeed, the operators generated by the Lanczos procedures are mutually orthonormal with respect to our chosen scalar product and $\mu_k = (\mathbf{B}|\mathbf{f}_k) \propto (\mathbf{f}_0|\mathbf{f}_k) = \delta_{0k}$, therefore only the $0$th moment can differ from 0.
The polynomials $L_k(\omega)$ produced by Lanczos' algorithm are defined by
\begin{align}
    \beta_{k+1}L_{k+1}(\omega) &= (\omega-\alpha_k)L_k(\omega) - \beta_k L_{k-1}(\omega)\\
    L_{-1}(\omega) &= 0\\
    L_0(\omega) &= 1
\end{align}
Those polynomials are orthonormal with respect to the spectral function $A_{CC}(\omega) = \lim_{\epsilon \rightarrow 0^+} \mathrm{Im}\left( \tfrac{-1}{\pi} G_{CC}(\omega +i\epsilon)\right)$
\begin{equation}
    \int_{-\infty}^\infty L_m(\omega)L_n(\omega)A_{CC}(\omega)d\omega = \delta_{mn}
\end{equation}
This observation provides us with an alternative way of computing off-diagonal Green function element within Lanczos' algorithm: we can compute their moments in the Lanczos-generated polynomial basis, yielding a decomposition of the off-diagonnal elements $ A_{\mathbf{BC}}(\omega)$ in terms of the polynomials that are orthogonal with respect to the targeted diagonal element $A_{\mathbf{CC}}(\omega)$.

In short, with this {\it continued fraction-polynomial hybrid} (CFPH) procedure, the diagonal Green function $G_{\mathbf{CC}}(z)$ is given by the continued fraction generated by Lanczos' algorithm and we have the coefficients $\mu_n = (\mathbf{B}|\mathbf{f}_n)$ for the off-diagonal Green function elements
\begin{equation}
   G_{\mathbf{BC}}(z) = \sum_n \mu_n \left( Q_n(z) + L_n(z){G_{\mathbf{CC}}(z)} \right),
   \label[type]{eq:Green_poly_hybrid}
\end{equation}
where the secondary polynomials $Q_n(\omega)$ are defined by 
\begin{align}
    \beta_{k+1}Q_{k+1}(z) &= (z-\alpha_k)Q_k(z) - \beta_kQ_{k-1}(z)\\
    Q_0(z) &= 0 \\
    Q_1(z) &= \frac{-L_1(z)}{z-\alpha_0}
\end{align}

This computation method has a significant advantage in comparison to either method taken alone.
Compared with the LCCF method for off-diagonal elements of the Green function, it is computationally cheaper.
Compared with the orthonormal polynomial computation method, there is no need to rescale the polynomials to fit the a priori unknown bandwidth of the Green function.
We speculate that moments computed on that basis converge exponentially, or faster.

For a generic $L$-site system, with no spatial nor time-reversal symmetry, the LCCF method would have to be performed $L^2$ times, requiring $L^2k$ actions of the Liouvillian on an operator and $2L^2k$ inner products, where $k$ is the number of iterations performed.
In contrast, the CFPH method would be performed $L$ times, requiring $Lk$ application of the Liouvillian and $(2+L(L-1))k$ inner products.
This is only $2k$ more inner products than the orthonormal polynomial method.

\subsection{Examples of scalar products}
\label[type]{subs:scalarprod}
The precise definition of the scalar product used determines the physical content of the response function the recursive methods computes.
Here we give examples of scalar products for a few common uses and demonstrate that they have all the properties required of inner products~\cite{jain_functional_2004}, namely positive definiteness, conjugate symmetry and linearity, along with the Hermiticity of the Liouvillian. 

The scalar products we consider here all have the form
\begin{equation}
    (\mathbf{B}|\mathbf{C}) = \frac{1}{\beta}\int_0^\beta d\lambda g(\lambda) \langle e^{\lambda \mathbf{H}}\mathbf{B}^\dagger e^{-\lambda \mathbf{H} } \mathbf{C}\rangle
    \label[type]{eq:general_scalar}
\end{equation}
where the angled brackets denote the thermodynamic average at inverse temperature $\beta$, $g(\lambda)$ is positive on $[0,\beta]$, $g(\beta-\lambda) = g(\lambda)$ and $\tfrac{1}{\beta}\int_0^\beta d\lambda g(\lambda)$ is some finite constant.
The thermodynamic average is defined as
\begin{equation}
    \langle \mathbf{B} \rangle = \frac{ \mathrm{Tr}\left[e^{-\beta \mathbf{H}} \mathbf{B} \right]}{\mathrm{Tr}\left[e^{-\beta \mathbf{H}}\right]}.
\end{equation}
Because the average in {Dirac's Hilbert space}\footnote{ By this we mean the usual Hilbert space of quantum mecanics, defined on wavefunction and operators} is both linear and conjugate symmetric, the linearity $(\mathbf{B}|\mathbf{C+D}) = (\mathbf{B}|\mathbf{C}) + (\mathbf{B}|\mathbf{D}) $ and conjugate symmetry $ (\mathbf{B}|\mathbf{C})^* = (\mathbf{C}|\mathbf{B}) $ are quite obvious.
But positivity $(\mathbf{B}|\mathbf{B}) \geq 0$ is less so.
Using completeness relations in terms of the eigenbasis of $\mathbf{H}$ we can write the scalar product as follows:
\begin{equation}
    (\mathbf{B}|\mathbf{B}) = \frac{1}{\beta}\int_0^\beta d\lambda g(\lambda) \sum_{nm} e^{(\lambda-\beta)E_n -\lambda E_m} | \langle m|\mathbf{B}|n \rangle |^2.
\end{equation}
In this form, we see that all the terms in the summation are positive.
Since $g(\lambda)$ is also positive, the integrand is positive everywhere in the integration domain.
Consequently, the formula is positive.

Finally, the Liouvillian is Hermitian with respect to this scalar product:
\begin{widetext}
\begin{equation}
    \begin{split}
        (\mathbf{B}|\mathcal{L}\mathbf{C}) &= \frac{1}{\beta}\int_0^\beta d\lambda g(\lambda) \langle e^{\lambda \mathbf{H}}\mathbf{B}^\dagger e^{-\lambda \mathbf{H} } [\mathbf{H},\mathbf{C}]\rangle \\
        &=\frac{1}{\beta \mathrm{Tr}[e^{-\beta \mathbf{H}}]}\int_0^\beta  d\lambda g(\lambda) 
        \mathrm{Tr}\left[ e^{(\lambda-\beta) \mathbf{H}}\mathbf{B}^\dagger  e^{-\lambda \mathbf{H} }\mathbf{H}\mathbf{C} 
         - e^{(\lambda-\beta)\mathbf{H}} \mathbf{B}^\dagger  e^{-\lambda \mathbf{H} }\mathbf{C}  \mathbf{H}\right] \\
        &=\frac{1}{\beta \mathrm{Tr}[e^{-\beta \mathbf{H}}]}\int_0^\beta  d\lambda g(\lambda) 
        \mathrm{Tr}\left[ e^{(\lambda-\beta) \mathbf{H}}\mathbf{B}^\dagger \mathbf{H} e^{-\lambda \mathbf{H} }\mathbf{C} 
         - e^{(\lambda-\beta) \mathbf{H}}\mathbf{H} \mathbf{B}^\dagger  e^{-\lambda \mathbf{H} }\mathbf{C} \right] \\
        &= \frac{1}{\beta}\int_0^\beta d\lambda g(\lambda) \langle e^{\lambda \mathbf{H}}\mathbf [\mathbf{B}^\dagger,\mathbf{H}] e^{-\lambda \mathbf{H} }\mathbf{C}\rangle \\
        &=\frac{1}{\beta}\int_0^\beta d\lambda g(\lambda) \langle e^{\lambda \mathbf{H}}\mathbf [\mathbf{H},\mathbf{B}]^\dagger e^{-\lambda \mathbf{H} }\mathbf{C}\rangle \\
        &= (\mathcal{L}\mathbf{B}|\mathbf{C})
    \end{split}
\end{equation} 
\end{widetext}
\subsection*{Fermionic Green function}
The practical definition of the many-body retarded Green function and spectral function for fermions in a system at equilibrium and as a function of real time are 
\begin{align}
    G_{rr'}^+(t) &= -i \langle \{ \mathbf{c}_r(t),\mathbf{c}^\dagger_{r'}(0) \}\rangle\theta(t) 
    \label[type]{eq:fermion_green}
    \\
    A_{rr'}^+(t) &= \langle \{ \mathbf{c}_r(t),\mathbf{c}^\dagger_{r'}(0) \}\rangle 
    \label[type]{eq:fermion_weight}
\end{align} 
where $\mathbf{c}^{(\dagger)}_r(t)$ is the annihilation (creation) operator for a fermion at position $r$ (and possibly other degrees of freedom) at time $t$ and the fermion operators obey the canonical anticommutation relation: $\{\mathbf{c}_r,\mathbf{c}^\dagger_{r'}\} = \delta_{rr'}\mathbf{I}$ and $\{\mathbf{c}_r,\mathbf{c}_{r'}\}= 0$.

This leads us to choose the following scalar product:
\begin{equation}
    (\mathbf{B}|\mathbf{C}) = \langle \{ \mathbf{C},\mathbf{B}^\dagger \}\rangle .
    \label[type]{eq:fermion_scalar_product}
\end{equation}
This scalar product corresponds to Eq.~\eqref{eq:general_scalar} with $g(\lambda) = \beta (\delta(\lambda) + \delta(\beta-\lambda))$.
Thus, it satisfies all the necessary properties.
\subsection*{Bosonic Green function}
The usual definition of the retarded  Green function and spectral function when dealing with bosons are
\begin{align}
    G_{rr'}^-(t) &= -i \langle [ \mathbf{b}_r(t),\mathbf{b}^\dagger_{r'}(0) ]\rangle\theta(t) 
    \label[type]{eq:boson_green}
    \\
    A_{rr'}^-(t) &= \langle [\mathbf{b}_r(t),\mathbf{b}^\dagger_{r'}(0) ]\rangle 
    \label[type]{eq:boson_weight}
\end{align}
where $\mathbf{b}^{(\dagger)}_r(t)$ is the annihilation (creation) operator for a boson at position $r$ (and possibly other degrees of freedom) at time $t$ and the boson operators obey the canonical commutation relation: $[\mathbf{b}_r,\mathbf{b}^\dagger_{r'}] = \delta_{rr'}\mathbf{I}$ and $[\mathbf{b}_r,\mathbf{b}_{r'}]= 0$.

This hints at $(\mathbf{B}|\mathbf{C}) = \langle [\mathbf{C},\mathbf{B}^\dagger ] \rangle$ for scalar product.
This scalar product is fiven by~\eqref{eq:general_scalar} with $g(\lambda) = \beta (\delta(\lambda) - \delta(\beta-\lambda))$.
This violates the positivity condition on $g(\lambda)$.
Therefore, this isn't a valid inner product for the recursive method.
The choice of a commutator instead of an anticommutator is a matter of convenience, and in this case, it is rather inconvenient.
So we advise using the anticommutator scalar product for bosonic Green function as well.
Indeed, it is straightforward to obtain the commutator spectral function Eq.~\eqref{eq:boson_weight} from the anticommutator spectral function Eq.~\eqref{eq:fermion_weight} using the detailed balance relation between two terms present in both functions.
We define $A^>_{rr'}(t) = \langle \mathbf{b}_r(t)\mathbf{b}^\dagger_{r'}\rangle$ and $ A^<_{rr'}(t) = \langle \mathbf{b}^\dagger_{r'}\mathbf{b}_{r}(t)\rangle $.
Using the trace cyclic property, we can show that
\begin{align}
    A^>_{rr'}(t-i\beta) =& \ A^<_{rr'}(t)\\
    \implies e^{-\beta\omega}A^>_{rr'}(\omega) =& \ A^<_{rr'}(\omega)
    \label[type]{eq:detailed_balance}.
\end{align}
Since $A^+(\omega) = A^>(\omega)+A^<(\omega)$ and $A^-(\omega) = A^>(\omega)-A^<(\omega)$, it follows from Eq.~\eqref{eq:detailed_balance} that
\begin{equation}
    A^-_{rr'}(\omega) = \mathrm{tanh}(\tfrac{\beta \omega}{2})A^+_{rr'}(\omega).
\end{equation} 
The corresponding Green function can be obtained using Eq.~\eqref{eq:Stieltjes_SW}, if needed.
Alternatively, one can compute each of the two terms of the commutator separately, by assigning to each of them its own inner product, similarly to what is done in the following section.
\subsection*{Zero temperature Hamiltonian limit}
A special case of the Liouvillian formalism as presented here is the recursive method at zero temperature in the Hamiltonian formalism. 
The Hamiltonian formalism has one very significant advantage: it only involves operator acting on states. 
The algebra of operators acting on states is a significantly smaller burden than that of operators and superoperators.
Using it when possible can lead to significantly faster numerical computations.

In the Hamiltonian formalism, the recursion method must be applied once per term in the spectral function~\cite{dionne_pyqcm_2023,viswanath_recursion_1994}.
To replicate this feature in the Liouvillian formalism, we must define two inner products, one for each of the two terms in the Green function:
\begin{align}
    (\mathbf{B}|\mathbf{C})^> &= \langle \C \B^\dagger \rangle\\
    (\mathbf{B}|\mathbf{C})^< &= \langle \B^\dagger \C \rangle.
\end{align}
Those scalar products correspond to Eq.~\eqref{eq:general_scalar} with $g^+(\lambda)=\beta \delta(\lambda)$ and $ g^-(\lambda) = \beta \delta(\beta-\lambda)$, respectively.
The fermionic Green function is then $ G_{rr'}(t) = \left[ (\mathbf{c}_{r'}|\mathbf{c}_{r}(t))^> +(\mathbf{c}_{r'}|\mathbf{c}_{r}(t))^<   \right]\theta(t)$.
The manipulations to be done are similar for both terms.
Taking the zero temperature limit, the Lanczos coefficient are 
\begin{align}
    \alpha^<_k =& \bra{\Omega} \mathbf{f}_{k,<}^{\dagger} (E_0-\mathbf{H}) \mathbf{f}_{k,<} \ket{\Omega}\\
    \beta^<_k =& \bra{\Omega} \mathbf{f}_{k,<}^{\dagger} (E_0-\mathbf{H}) \mathbf{f}_{k-1,<} \ket{\Omega}\\
    \alpha^>_k =& \bra{\Omega} \mathbf{f}_{k,>} (\mathbf{H}-E_0) \mathbf{f}_{k,>}^{\dagger} \ket{\Omega}\\
    \beta^>_k =& \bra{\Omega} \mathbf{f}_{k-1,>} (\mathbf{H}-E_0) \mathbf{f}_{k,>}^{\dagger} \ket{\Omega}
\end{align}
where $\ket{\Omega}$ is the non-degenerate ground state and $\mathbf{f}_{0,\lessgtr} = \mathbf{c}_r$.
We observe that we can express both the Lanczos coefficient Eqs.~(\ref{eq:Lanczos_alpha},\ref{eq:Lanczos_beta}) and the recursion relation Eq.~\eqref{eq:LLanczos_recursion} in terms of the states $\ket{\psi_k^\lessgtr} = \mathbf{f}_{k,\lessgtr} \ket{\Omega}$.
Indeed, by applying the recursion relation to the ground state we obtain 
\begin{align}
    \beta_{k+1}^< \ket{\psi^<_{k+1}} = ( -(H-E_0) -\alpha^<_k)\ket{\psi^<_k} -\beta^<_k\ket{\psi^<_{k-1}}
\end{align}
and
\begin{align}
    \beta_{k+1}^> \ket{\psi^>_{k+1}} = (H-E_0-\alpha^>_k)\ket{\psi^>_k} -\beta^>_k\ket{\psi^>_{k-1}}.
\end{align}
These two recursion relations on states are equivalent to what one finds when applying the Lanczos method to the computation of Green functions within the Hamiltonian formalism~\cite{dionne_pyqcm_2023,premont-foley_reseaux_2020,dagotto_correlated_1994}.

\section{Model}
To illustrate the recursion method, we apply it to the study the fermionic Hubbard model on a $2\times2$ plaquette.
Finite realizations of the Hubbard model such as this are relevant to cluster methods~\cite{maier_quantum_2005,seki_variational_2018,aichhorn_variational_2006,potthoff_making_2014,dionne_pyqcm_2023,dahnken_variational_2004,senechal_competition_2005}.
This family of methods yields an approximate Green function of a lattice model from that of a plaquette.
With methods of that nature, only quantities expressible in terms of the Green function can be computed for the infinite lattice.

The Hubbard Hamiltonian is 
\begin{equation}
\begin{split}
    H =& \ t\sum_{\langle r,r'\rangle \sigma} \left(\mathbf{c}_{r'\sigma}^\dagger \mathbf{c}_{r\sigma} + \mathbf{c}_{r\sigma}^\dagger \mathbf{c}_{r'\sigma}\right)\\
    &+ U\sum_r \mathbf{c}^\dagger_{r\uparrow}\mathbf{c}^\dagger_{r\downarrow}\mathbf{c}_{r\downarrow} \mathbf{c}_{r\uparrow}
    - \mu \sum_{r\sigma} \mathbf{c}_{r\sigma}^\dagger \mathbf{c}_{r\sigma} 
\end{split}
\end{equation}
where $r$ is a position on the plaquette, $\sigma \in \{\uparrow,\downarrow\}$ is the spin index, $t$ is the hopping amplitude, $U$ the on-site Hubbard interaction and $\mu$ the chemical potential.
$t\equiv 1$ defines the energy scale.
We will consider the intermediate coupling regime $U=4$ at half-filling $\mu = \tfrac{U}{2}$.
In this particular realization, all four sites are equivalent, and the model is spin symmetric.
Due to the model finite size, no phase transition can happen and no symmetry can be broken spontaneously.
Thus, we make use of all available symmetries: it is enough to compute the Green function elements at $r=(0,0)$ and $r'\in[(0,0),(0,1),(1,1)]$ and only for one spin polarisation to reconstruct the Green function in its entirety. 

The complete Hilbert space for this model has size $4^4=256$.
Consequently, commonly available linear algebra implementations are good enough to compute results in very little time for this particular model.
Here we use numpy to perform the numerical computations.
For this problem, we perform full diagonalization of the plaquette Hamiltonian $\mathbf{H}$ and compute the density matrix 
\begin{equation}
\rho = \frac{\mathrm{exp}(-\beta \mathbf{H})}{\mathrm{Tr}(\mathrm{exp}(-\beta \mathbf{H}))}.
\end{equation}
The scalar product ~\eqref{eq:fermion_scalar_product} is used to compute the retarded Green function ~\eqref{eq:fermion_green}.
Much more care and efforts would be needed to tackle larger finite models with this category of methods.
Indeed, the cost of manipulating operators in a given Hilbert space is comparable to that of manipulating a state in a system twice the size and full diagonalization of the problem Hamiltonian is not practical for only slightly larger problems.

We perform a temperature-dependence study of the plaquette by computing its local density of states and heat capacity from Green function using exclusively continued fractions.
This demonstrates the capability of this method to accurately compute various equilibrium quantities, as the system's free energy can be reliably determined if the heat capacity can be accurately computed.
Additionally, we compare the off-diagonal elements of Green function obtained by LCCF (Eqs.~(\ref{eq:LCCFR},\ref{eq:LCCFI})) with that from the CFPH procedure ~\eqref{eq:Green_poly_hybrid}.
This comparison shows that the CFPH procedure reproduces the same spectral function as the LCCF to high accuracy, with significantly fewer computations.

\subsection*{Results}
\begin{figure*}[htb]
    \includegraphics[width = 0.75\textwidth]{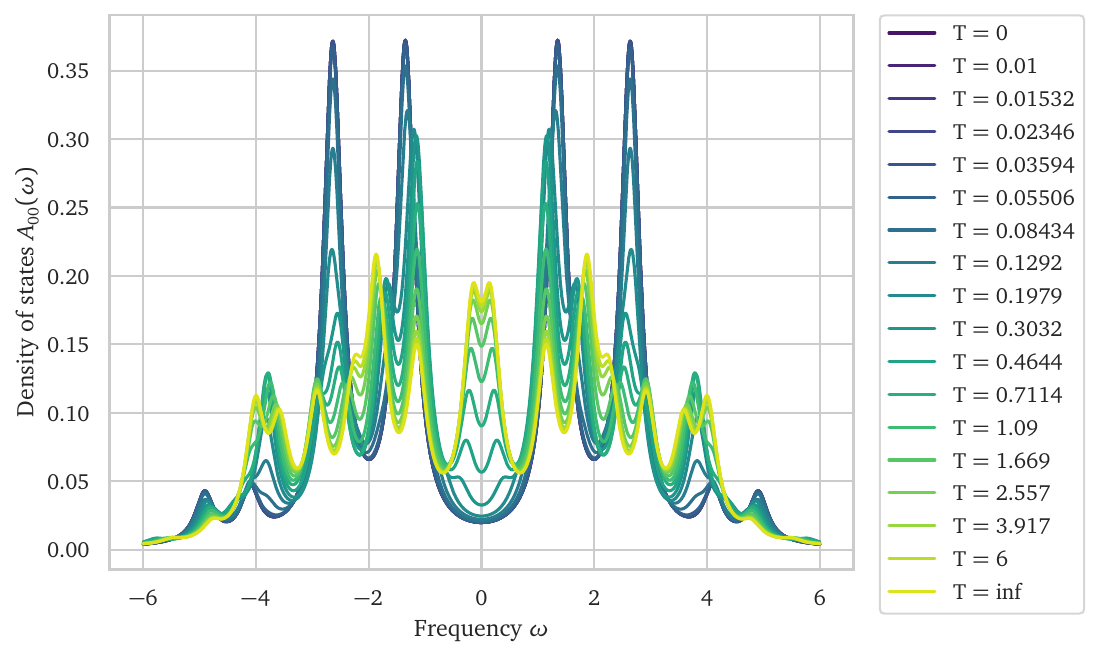}
    \caption{Local density of state as a function of frequency at temperatures ranging from 0 to infinity with Lorentzian broadening $\eta=0.2$ and at half-filling with $U=4$.
    The density of states is computed by taking the imaginary part of the local Green function at a given temperature.
    The Green function is represented in the form of a continued fraction with at most $k=40$ floors.
    At low temperatures, fewer than $40$ floors are needed.}
    \label[type]{fig:Ldos}
\end{figure*}

In Fig.~\ref{fig:Ldos}, we see the evolution of the local density of states with temperature.
The local density of states as shown in the figure is 
\begin{equation}
    D(\omega) = A_{00}(\omega) = \frac{-1}{\pi}\mathrm{Im}\left( G(w+0.2i)\right).
\end{equation}
The limit of real frequencies is not taken, leading to a Lorentzian broadening of the Dirac deltas that would otherwise be present at real frequencies.
We observe that most of the temperature driven changes occur before $T=1$ and that by $T\approx 2.5$ the infinite temperature limit is almost reached.
At $T\approx 0.3$ a significant amount of spectral function is visible in the Mott gap.
This is a hint of the temperature driven metal-insulator transition in the lattice limit.
As temperature rises, this weight within the gap keeps getting larger, a clear sign that the metallic phase is favoured by higher temperature.

As temperature rises from $0$, the number of peaks visible in the spectrum increases significantly.
At zero temperature, the exact density of state has sixteen poles, many of which have weights so small that they are not visible.
Consequently, the Lanczos algorithm naturally stops after 16 iterations.
As temperature rises, the number of poles in the exact solution rises quickly and before $T=0.055$ it becomes limited to $40$, the maximum number of iterations of the Lanczos algorithm we allowed for those computations.
For this reason, the total number of poles present in the solutions is fixed and their positions lose their exact correspondence to excited states eigenvalues.
At $T=0.055$, most of the temperature-driven changes to the spectrum have yet to happen: the changes we observe happen through a reorganization of the available poles.

\begin{figure}[h]
    \includegraphics[width = 0.95\columnwidth]{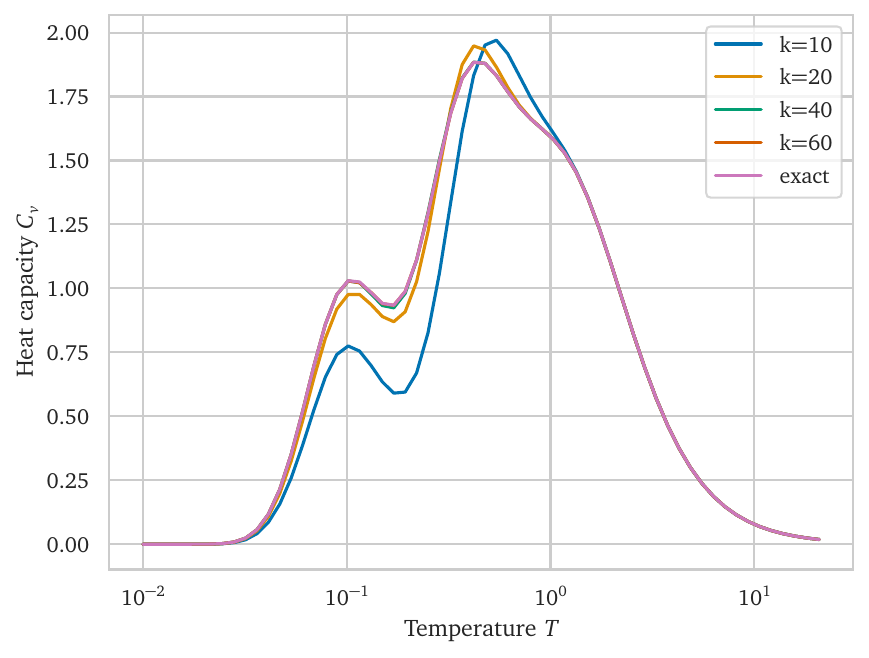}
    \caption{Heat capacity of the $2\times 2$ Hubbard plaquette as a function of temperature at half-filling with interaction $U=4$.
The different $k$ values are the maximum number of floors in the continued fractions, the exact result is computed directly from the density matrix.}
    \label{fig:CvT}
\end{figure}

\begin{figure}[h]
    \includegraphics[width = 0.95\columnwidth]{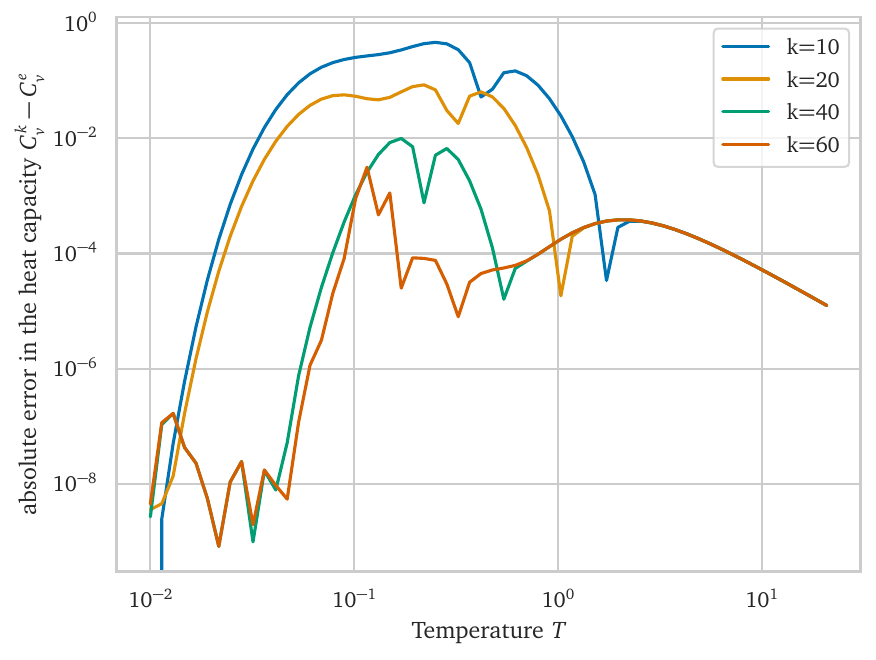}
    \caption{Error in the heat capacity with respect to the numerically exact computation.
}
    \label[type]{fig:DCvT}
\end{figure}

In practice, the limited number of iteration needs not limit the accuracy of computations, thermodynamic quantities show a good convergence rate with the maximum number of poles allowed. 
Figures~\ref{fig:CvT} and \ref{fig:DCvT} illustrate this quite neatly.
In those we compute the constant volume heat capacity as a function of temperature using the Green function and compare the resulting estimate with the numerically exact value. 
The exact value is obtained by computing
\begin{equation}
    C_V(T) = \frac{1}{T^2}\left( \langle \mathbf{H}^2 \rangle - \langle \mathbf{H}\rangle^2 \right)
\end{equation}
using the density matrix.
The Green function estimate is obtained by numerical differentiation of the total energy with respect to inverse temperature:
\begin{align}
    E(T) &= \frac{1}{2}\mathrm{Tr}\left[\oint_C dz (z+H_0)f_T(z)G_T(z)\right]\\
    C_V(T) &= \left( \frac{\delta E(T)}{\delta T} \right) _V = \frac{-1}{T^2}\left(\frac{\delta E(T)}{\delta \beta} \right)_V
\end{align}
where $f_T(z)$ is the Fermi-Dirac distribution at temperature $T$, $H_0$ is the one-body Hamiltonian matrix and the closed integration path $C$ contains all the poles of the Green function $G_T(z)$ and none of the poles of Fermi-Dirac's distribution.
The numerical derivative is done with the symmetric finite difference formula
\begin{equation}
    \frac{\delta E(\beta)}{\delta \beta} = \frac{E(\beta+\Delta) - E(\beta - \Delta)}{2 \Delta} + O(\Delta^2),
\end{equation}
 with an inverse temperature variation $\Delta = 0.01$.

In Fig.~\ref{fig:CvT}, we observe a good qualitative agreement with the exact result with as few as $10$ floors in the continued fractions used.
Indeed, all the expected features are present in roughly the correct positions, with similar relative amplitudes.
With only $20$ floors in the continued fractions, the agreement is nearly quantitative; the absolute error is less than $0.1$, as can be seen in Fig.~\ref{fig:DCvT}. 
With $40$ and $60$ floors in the continued fractions, quantitative agreement is reached for all values. 
Note that at temperatures above $1$, the heat capacity computations gradually become independent of the number of poles; the computations' accuracy is limited by the finite difference formula we used, which has error $O(10^{-4})$.
Since this approach to computing heat capacity depends on an accurate computation of the total energy, it means that the free energy can also be computed accurately.
Indeed, the free energy is~\cite{sup_mat}  
\begin{equation}
\begin{split}
    F(T) &= E(T) - TS(T) = E(T) - T\int_0^T dT_0 \frac{C_V(T_0)}{T_0} \\
    &= E(0) - T\int_0^T dT_0 \frac{E(T_0)-E(0)}{T_0^2}.
\end{split}
\end{equation}
Consequently, any quantity that can be expressed as a derivative of the free energy can be computed with an error comparable to that of the heat capacity.

\begin{figure}[ht]
    \includegraphics[width = 0.95\columnwidth]{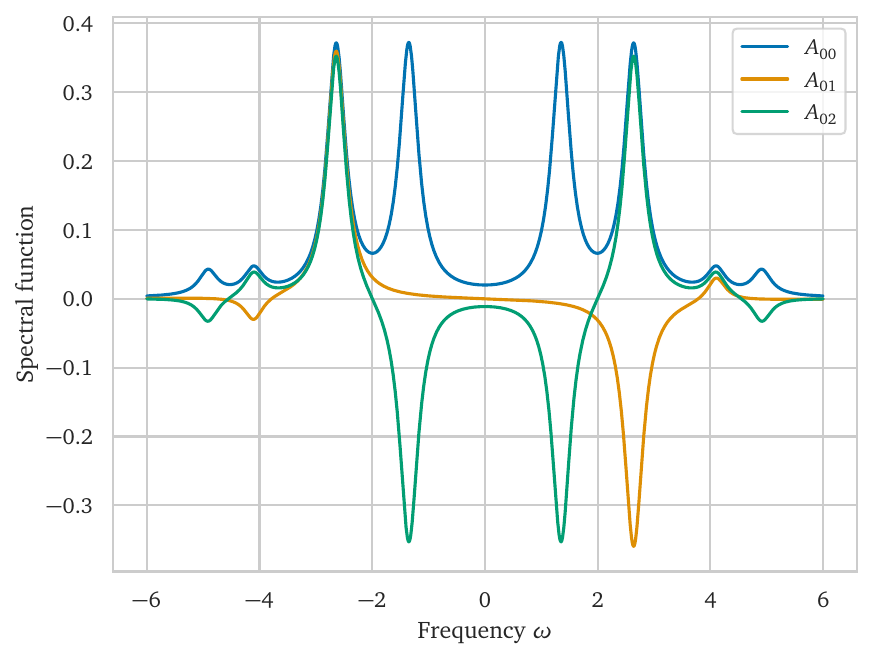}
    \caption{Spectral function, on-site ($A_{00}$), first neighbours ($A_{01}$) and second neighbours ($A_{02}$) at inverse temperature $\beta=80$ with Lorentzian broadening $\eta=0.2$.}
    \label[type]{fig:SW_poly}
\end{figure}

\begin{figure}[hb]
    \includegraphics[width = 0.95\columnwidth]{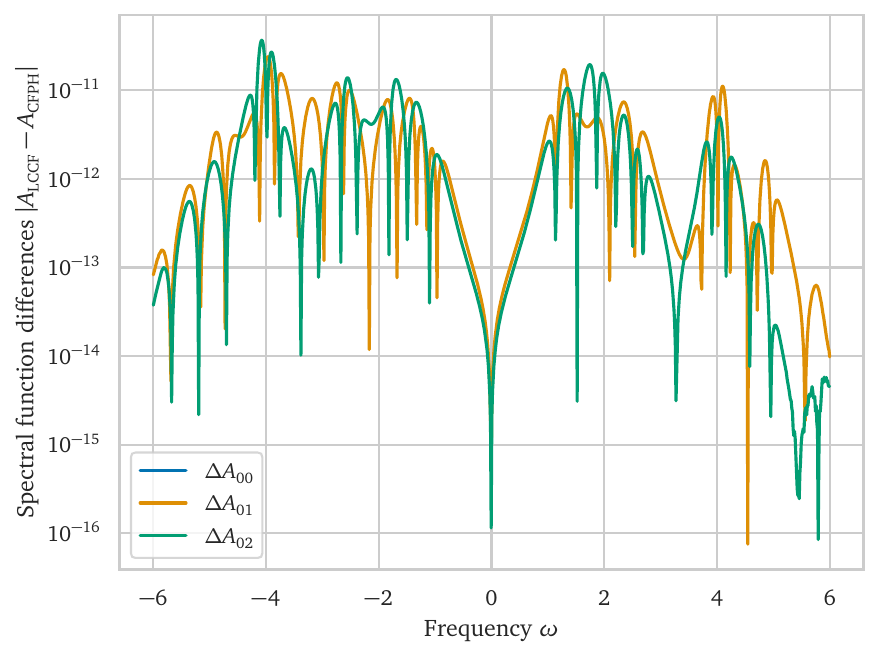}
    \caption{Absolute difference in the spectral function computed with the CFPH and the LCCF method.}
    \label[type]{fig:err_SW_poly}
\end{figure}

In Fig.~\ref{fig:SW_poly} we show the three irreducible contribution to the entire spectral function, computed with the CFPH method.
The results of the LCCF approach to the computation of off-diagonal elements are not shown because there would be no visible difference. 
Indeed, the resulting spectral functions are numerically equivalent, as shown in Fig.~\ref{fig:err_SW_poly}.
We observe there that the most pronounced difference is less than $10^{-10}$. 

Computing all three irreducible elements of the Green function of this particular model can be done in a single recursive procedure, needing only one extra inner product computation per iteration per off-diagonal element, bringing the total number of operations to $k$ application of the Liouvillian and $(2+o)k$ inner product computations, where $k\leq40$ is the number of iterations of the recursion process and $o=2$ is the number of off-diagonal Green function elements to compute.
This is in contrast to $(o+1)k$ application of the Liouvillian and $2ok$ inner products of the LCCF method.

\section{Conclusion} 
In conclusion, we have presented a comprehensive and simplified framework for the recursion method. 
This framework allows for the computation of response functions for a wide range of quantum operators within the Liouvillian formalism, including non-Hermitian operators. 
Notably, we have shown that the Hamiltonian formalism of the recursion method is a special case of the Liouvillian recursion method. 
 
While we have focused on a toy problem, it is important to note that the recursion method has already been successfully applied to problems of interest in condensed matter;
Its ability to accurately compute response functions and extract valuable information for quantum many-body systems has been demonstrated in previous studies~\cite{viswanath_recursion_1994,lee_derivation_1983,lee_orthogonalization_1982,lee_solutions_1982,mori_continued-fraction_1965}.
The generality and flexibility of the method make it a good tool for investigating a wide range of strongly correlated systems.

However, further efforts are required to design and implement efficient algorithms and techniques specifically tailored to larger and more realistic systems. 
Tensor network methods~\cite{orus_tensor_2019}, for instance, show promise in addressing the computational challenges associated with larger systems.
With tensor networks, time and memory needs are nominally polynomial.
These methods have polynomial time and memory requirements, but their accuracy depends on the choice of a parameter, the bond dimension.
The precise effect of that parameter on accuracy is problem dependent and its contribution to computational complexity depends on network topology. 
Consequently, constant precision complexity is often discovered through empirical means. 
For limited order development, we find an alternative in variational representations~\cite{sorella_wave_2005} of the thermodynamic state along with a sparse representation of the operator generated by the recursion relations. 
In addition to the complexity of producing the equilibrium state, the worst-case time and memory needs are of order $O(N^k)$ where $N$ is the number of terms in the sparse representation of the Hamiltonian and $k$ is the order to which the recursive procedure is carried out.
Variational solutions on near-term quantum computers~\cite{zoufal_variational_2021,tilly_variational_2022} are of particular interest as a potential applications of recursion method.

In summary, the Liouvillian framework for recursion methods is a versatile approach for the computingh response functions in quantum many-body systems.
By further refining its algorithms and combining it with complementary techniques, we are confident that complex quantum materials can be better understood.

\begin{acknowledgments}
Thanks to professors Maxime Charlebois, André-Marie Tremblay and David Sénéchal for insightful discussions and encouragement.
Thanks to my colleagues Jean-Fréderic Laprade, Maxime Dion, Marco Armenta, Tania Belabas, Ghislain Lefebvre and Christian Sara-Bournet for their help, support and encouragement.
Finally, thanks to Jérôme Leblanc for his careful reading of the many drafts of this document.
This work has been supported by the Canada First Research Excellence Fund and the Ministère de l'économie, de l'innovation et de l'énergie du Québec.
\end{acknowledgments}
 
\bibliography{LL_article.bib}

\pagebreak

\end{document}


\title{Supplementatry materials: Liouvillian recursion method for the electronic Green's function}

\author{A. Foley}
\email[Corresponding author: ]{Alexandre.Foley@usherbrooke.ca}
\affiliation{AlgoLab Quantique, Institut Quantique, Universit\'e de Sherbrooke, Sherbrooke, QC, Canada}
\date{\today}

\begin{abstract}
    In this document, we give a proof of the identity for the Free Energy $ F = E(0) - T\int_0^TdT_0 \tfrac{E(T_0)-E(0)}{T_0^2}$ and for the continued fraction representation of the Green funciton.
    The first was found without proof on physics exchange~\cite{phys_ex_FG}. It's likely that it can be found in some textbook.
    The second has been demonstrated in the literature, but this particular proof differ from the commonly cited one by Dagotto~\cite{dagotto_correlated_1994} in ways we believe are interesting and enlightening.
    The author of this work first published this second proof in french his thesis~\cite{premont-foley_reseaux_2020}.
\end{abstract}

\maketitle
 
\section{Proof of $ F(T) = E(0) - T\int_0^TdT_0 \tfrac{E(T_0)-E(0)}{T_0^2}$} 

Our starting point is
\begin{equation}
    F(T) = E(T) - TS(T)
\end{equation}
where $F(T)$ is the free energy, $E(T)$ is the total energy, $S(T)$ is the entropy and $T$ is the temperature.
It is well known that 
\begin{equation}
    S(T) = \int_0^T dT' \frac{C_v(T')}{T'}.
\end{equation}
When all things but the temperature are kept constant $C_v(T) = \tfrac{d E(T)}{d T}$.
Therefore,
\begin{equation}
    S(T) = \int_0^T \frac{dT'}{T'} \frac{d E(T')}{d T'}.
    \label[type]{eq:int_F}
\end{equation}
Integrating by part, we obtain
\begin{equation}
    S(T) = \int_0^T \frac{dT'}{T'} \frac{d E(T')}{d T'} = \lim_{T_1 \rightarrow 0 } \left[\frac{E(T_0)}{T_0}\right]_{T_1}^T + \int_{T_1}^TdT_0 \frac{E(T_0)}{T_0^2}
\end{equation}
Because the total energy converge exponentially to its zero temperature limit, we can affirm we're dealing with an $\infty-\infty$ indeterminate form.
Taking out of the limit the single $T_1$ independent term, we get:
\begin{equation}
    S(T) = \frac{E(T)}{T} - \lim_{T_1\rightarrow 0}\left( \frac{E(T_1)}{T_1} - \int_{T_1}^TdT_0 \frac{E(T_0)}{T_0^2} \right)
\end{equation}
We proceed to add $0=\frac{E(0)}{T_1} - \frac{E(0)}{T_1}$ inside the limit to zero temperature:
\begin{equation}
\begin{split}
    S(T) &= \frac{E(T)}{T} - \lim_{T_1\rightarrow 0}\left( \frac{E(T_1)}{T_1} - \frac{E(0)}{T_1} + \frac{E(0)}{T_1} - \int_{T_1}^TdT_0 \frac{E(T_0)}{T_0^2} \right)\\
    &= \frac{E(T)}{T} - \lim_{T_1\rightarrow 0}\left( \frac{E(T_1) - E((0))}{T_1} \right)  + \lim_{T_1\rightarrow 0} \left( \int_{T_1}^T \frac{E(T_0)}{T_0^2}dT_0 - \frac{E(0)}{T_1}\right)
\end{split}
\end{equation}
The leftmost limit can be evaluated using l'Hospital rule.
\begin{equation}
\begin{split}
    S(T) &= \frac{E(T)}{T} - \lim_{T_1\rightarrow 0}\left( \frac{d E(T_1)}{d T_1} \right)  + \lim_{T_1\rightarrow 0} \left( \int_{T_1}^T \frac{E(T_0)}{T_0^2}dT_0 - \frac{E(0)}{T_1}\right) 
\end{split}
\end{equation}
From general thermodynamical considerations $\lim_{T_1 \rightarrow 0}\frac{d E(T_1)}{d T_1} = \lim_{T_1 \rightarrow 0} C_v(T_1) = 0$.
Using the fact that $ \tfrac{1}{T_1} = \int_{T_1}^T \tfrac{dT_1}{T_1^2} + \tfrac{1}{T}$, we obtain
\begin{equation}
\begin{split}
    S(T) &= \frac{E(T)}{T} + \lim_{T_1\rightarrow 0} \left(  \int_{T_1}^T \frac{E(T_0)-E(0)}{T_0^2}dT_0 - \frac{E(0)}{T}\right)\\
    &= \frac{E(T)-E(0)}{T} + \int_{0}^T \frac{E(T_0)-E(0)}{T_0^2}dT_0
\end{split}
\end{equation}
Substituting this expression for the entropy into the free energy and simplifying, we obtain the sought identity:
\begin{equation}
    F(T) = E(0) - T\int_0^TdT_0 \tfrac{E(T_0)-E(0)}{T_0^2}.
\end{equation}
QED.
\section{continued fraction representation of the Green's function}
We have to demonstrate that the topleftmost element of the inverse of $\omega-T$ is a continued fraction:
\begin{equation}
    [\omega-T]^{-1}_{(0,0)} = \cfrac{1}{\omega-\alpha_0-\cfrac{\beta_1^2}{\omega - \alpha_1 - \cfrac{\beta_2^2}{\ddots}}}
\end{equation}
Where $T$ is the tridiagonal matrix: 
\begin{equation}
T = \left(\begin{array}{cccccc}{\alpha_{0}} & {\beta_{1}} & {} & {} & {} & {0} \\ {\beta_{1}} & {\alpha_{1}} & {\beta_{2}} & {} & {} \\ {} & {\beta_{2}} & {\alpha_{2}} & {\ddots} & {} \\ {} & {} & {\ddots} & {\ddots} & {\beta_{m-2}} \\ {} & {} & {} & {\beta_{m-2}} & {\alpha_{m-2}} & {\beta_{m-1}} \\ {0} & {} & {} & {} & {\beta_{m-1}} & {\alpha_{m-1}}\end{array}\right)\label{eq:KryHam}
\end{equation}

We proceed by first defining the submatrices of $T$:
\begin{align}
T_i &= \left(\begin{array}{cccccc}{
	\alpha_{i}} & {\beta_{i+1}} & {} & {} & {} & {0} \\
{\beta_{i+1}} & {\alpha_{i+1}} & {\beta_{i+2}} & {} & {} \\ {} & {\beta_{i+2}} & {\alpha_{i+2}} & {\ddots} & {} \\ {} & {} & {\ddots} & {\ddots} & {\beta_{m-2}} \\
{} & {} & {} & {\beta_{m-2}} & {\alpha_{m-2}} & {\beta_{m-1}} \\
{0} & {} & {} & {} & {\beta_{m-1}} & {\alpha_{m-1}}\end{array}\right)_{(m-i)\times(m-i)} 
\end{align}
and the vectors
\begin{align}
B_i &= \left(\begin{array}{cccc} \beta_i & 0 & \dots & 0 \end{array} \right)_{1\times(m-i)} \textrm{.}
\end{align}
Note that $T_0 \equiv T$. The matrices $T_i$ can be recurently described in a block-matrix notation: 
\begin{equation}
	T_i =  \left( \begin{array}{cc}
		\alpha_i & B_{i+1} \\
		B^\dagger_{i+1} & T_{i+1}
	\end{array}\right) \textrm{.}
\end{equation}
Armed with this matrix by block expression, we can express the inversion problem in a neat manner:
\begin{equation}
	\left( \begin{array}{cc}
		z-\alpha_i & B_i \\
		B^\dagger_i & z-H_{i+1}
	\end{array}\right)
	\cdot
	\left( \begin{array}{cc}
		g_i & X_i \\
		X^\dagger_i & Z_i
	\end{array}\right)
	=
	\left( \begin{array}{cc}
		1 & 0 \\
		0 & \mathbb{1}
	\end{array}\right);
\end{equation}
and solve for the topleft element of the inverse matrix $g_i$, we find
\begin{equation}\label{eq:girec}
	g_i = \frac{1}{z-\alpha_i - B_{i+1}[z-T_{i+1}]^{-1}B^\dagger_{i+1}}.
\end{equation}
Using the fact that only the $(0,0)$ elements of the $B_i$ is nonzero, we rewrite the preceding as:
\begin{equation}
\begin{split}
	g_i &= \frac{1}{z-\alpha_i - \beta_{i+1}^2[z-T_{i+1}]^{-1}_{(0,0)} } \\
     &= \frac{1}{z-\alpha_i - \beta_{i+1}^2g_{i+1}}
\end{split}
\end{equation}
Therefore,
\begin{align}
    [\omega-T]^{-1}_{(0,0)} &= g_0\\
    &=\cfrac{1}{\omega-\alpha_0 - \beta^2_1 g_1}\\
    &=\cfrac{1}{\omega-\alpha_0 - \cfrac{\beta_1^2}{\omega-\alpha_1-\beta_2^2 g_2}}\\
    &=\cfrac{1}{\omega-\alpha_0-\cfrac{\beta_1^2}{\omega - \alpha_1 - \cfrac{\beta_2^2}{\ddots}}},
\end{align}
QED.

\bibliography{LL_article.bib}